\documentclass[aps,prl,reprint,amsmath,amssymb,superscriptaddress]{revtex4-1}

\usepackage{graphicx}
\usepackage{bm} 
\usepackage{verbatim}
\usepackage{hyperref}

\begin{document}

\title{\large Competition between Kondo screening and quantum Hall edge reconstruction}

\author{A. W. Heine}
\email{e-mail: heine@nano.uni-hannover.de}
\author{D. Tutuc}
\affiliation{Institut f\"ur Festk\"orperphysik, Leibniz
Universit\"at Hannover, Appelstr. 2, 30167 Hannover, Germany}
\author{G. Zwicknagl}
\affiliation{Institut f\"ur Mathematische
Physik, Technische Universit\"at Braunschweig, Mendelssohnstr. 3, 38106
Braunschweig, Germany}
\author{R.~J. Haug}
\affiliation{Institut f\"ur Festk\"orperphysik, Leibniz
Universit\"at Hannover, Appelstr. 2, 30167 Hannover, Germany}

\date{\today}


\begin{abstract}
We report on a Kondo correlated quantum dot connected to two-dimensional leads where we demonstrate the renormalization of the g-factor  in the pure Zeeman case i.e, for magnetic fields parallel to the plane of the quantum dot. For the same system we study the influence of orbital effects by investigating the quantum Hall regime i.e. a perpendicular magnetic field is applied. In this case  an unusual behaviour of the suppression of the Kondo effect  and of the split zero-bias anomaly is observed. The splitting decreases with magnetic field and shows discontinuous changes which are attributed to the intricate interplay between  Kondo screening and the quantum Hall edge structure originating from electrostatic screening.  This edge structure made up of compressible and incompressible stripes strongly affects the Kondo temperature of the quantum dot and thereby influences the renormalized g-factor.
\end{abstract}

\maketitle

\begin{figure}
\includegraphics[scale=0.5]{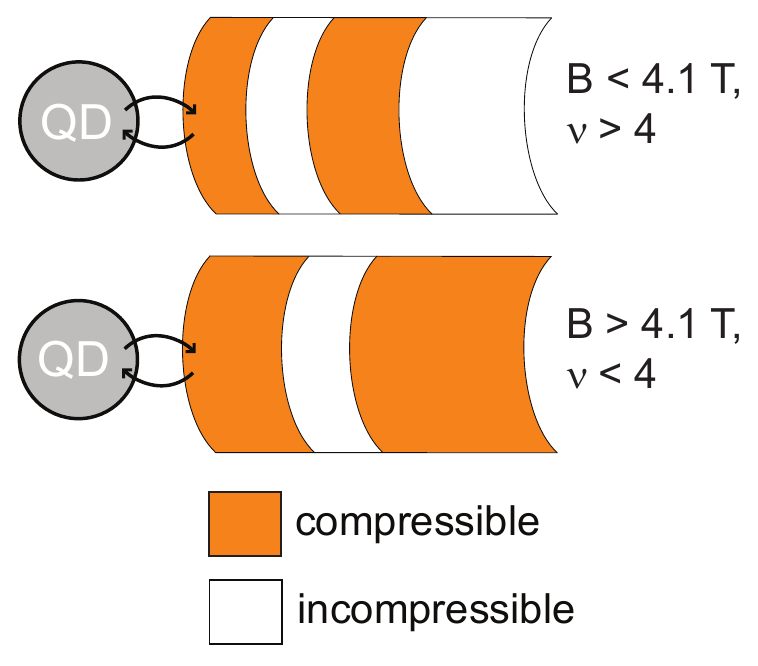}
\caption{Sketch of the edge channel structure near the QD for non-integer filling factors immediate above $\nu=4$ (top) and below $\nu=4$ (bottom), respectively.}
\label{edgechannels}
\end{figure}

The Kondo effect \cite{Hewson1993} is one of the most fascinating many-body correlation effects. It was first discovered in metals with magnetic impurities and was attributed to the screening of these local magnetic moments by conduction electrons. In Coulomb-blockaded quantum dots (QDs) acting as highly tunable single magnetic impurities the Kondo effect manifests itself as a zero-bias anomaly  (ZBA) in the enhanced conductance  \cite{Goldhaber-Gordon1998a,Goldhaber-Gordon1998,Cronenwett1998,Schmid1998,Wiel2000}. ZBAs appear quite often in transport experiments as a result of interaction effects, e.g. the claim of the observation of Majorana fermions was related to a ZBA \cite{Mourik2012,Lutchyn2010,Oreg2010}.  The origin of such a ZBA can be identified only by the detailed understanding of the influence of magnetic fields  as in the case of the Majorana fermions. But the influence of magnetic fields on interaction and correlation effects can be quite intricate.  In general the ZBA of the Kondo effect is suppressed by magnetic field. The accepted physical picture is that an external magnetic field polarizes the magnetic impurity removing the spin degeneracy which is the prerequisite for the appearance of the small energy scale. The low-energy excitations giving rise to the Kondo resonance can be described in terms of fermionic strongly renormalized "heavy" spin-1/2 quasiparticles.  In a magnetic field the energies of spin-1/2 quasiparticles with spin $\sigma=\pm$ will be shifted  by the Zeeman energy $\tilde{b}_\sigma=\frac{1}{2}\tilde{g}\sigma\mu_B B$ where $B$ and $\mu_B$ are the magnetic field and the Bohr magneton, respectively. The effective g-factor, $\tilde{g}$, whose bare value is $g=0.44$ for GaAs is renormalized by the Kondo effect \cite{Hewson2006,Edwards2011}. It assumes twice its bare value in the limit of low external fields as predicted by Wilson \cite{Wilson1975}. The central focus of the present study is the evolution of the Zeeman splitting with magnetic field. The system under consideration is a Kondo-correlated QD connected to two-dimensional leads. Of particular interest is a discontinuous
change in the Zeeman splitting of the ZBA observed upon tuning the filling factor of the leads in the quantum Hall regime through an integer
value. Our conjecture is that the abrupt changes reflect the reconstruction of the quantum Hall edge sketched schematically in Fig. \ref{edgechannels}.

\begin{figure}
\includegraphics[scale=0.5]{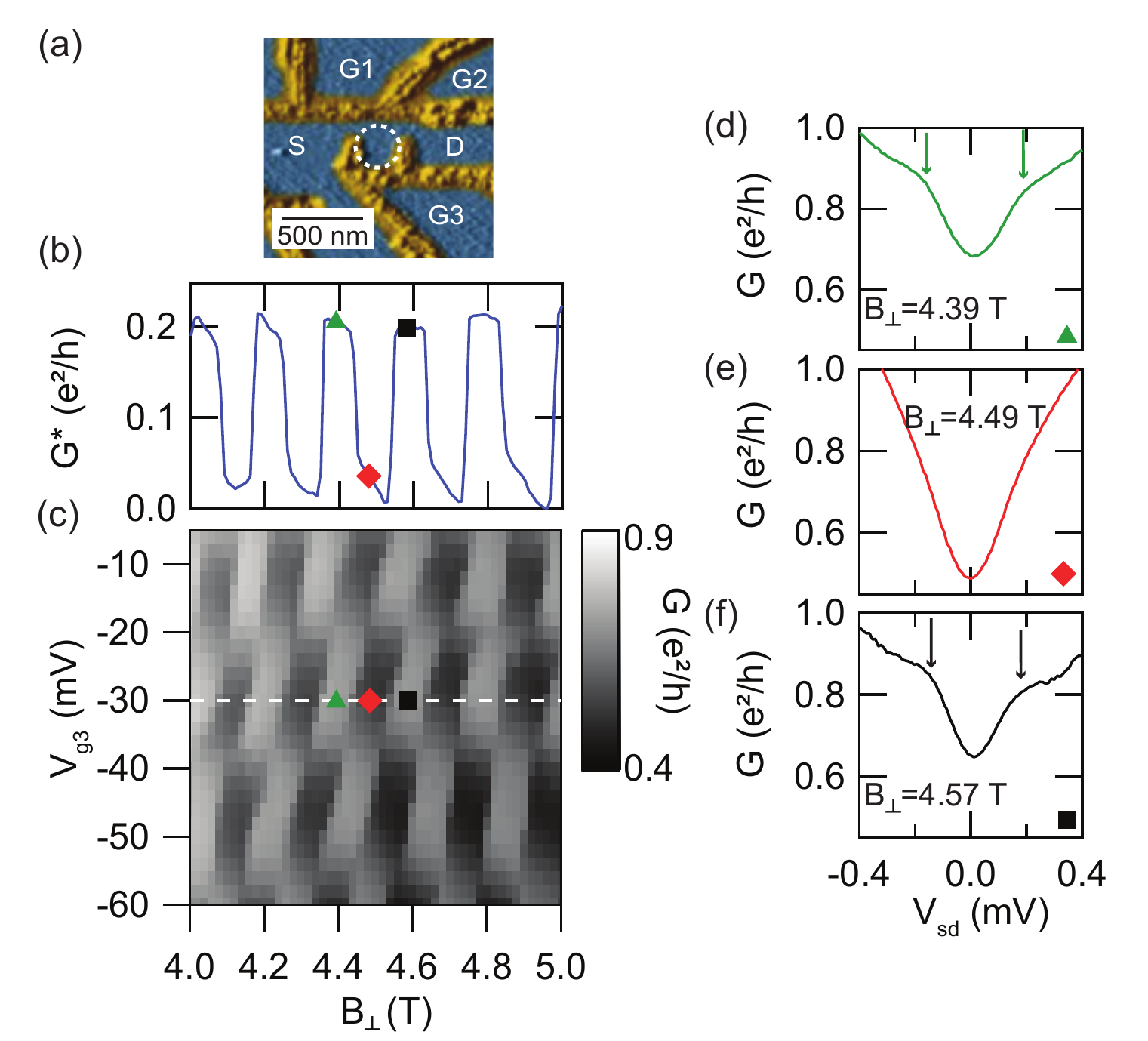}
\caption{(a) An AFM image of the sample. The oxide lines that define the structure are marked in yellow. (b) Difference $G^*$ in the differential conductance between Kondo enhanced and Coulomb blocked regions as function of magnetic field along the dashed line in (c). The background was subtracted. (c) Differential conductance through the QD as function of perpendicular magnetic field $B_\perp$ and gate voltage $V_\text{G3}$. (d-f) Bias dependence of $G$ for fixed values of $B_\perp$ referring to points in (b) and (c), which are marked by respective symbols. Shoulders are indicated by arrows.}
\label{perpendicular1}
\end{figure}

The experimental data were obtained on a QD fabricated in a two-dimensional electron gas (2DEG) 37~nm below the surface. The electron density of $n_e=3.95 \cdot 10^{15}$~m$^{-2}$ and the electron mobility of $\mu_e=65.6$~m$^2$/Vs of the 2DEG was determined at liquid Helium temperature. A QD is formed by oxide lines produced by local anodic oxidation\cite{Held1998,Keyser2000}. Its diameter is about 140~nm and it is coupled to source (S) and drain (D) via tunnel barriers. Three in-plane gates (G1, G2, and G3) are used to control the tunnel coupling to the leads and the energy levels on the dot \cite{Tutuc2011}. An atomic force microscope (AFM) image of the sample is shown in Fig. \ref{perpendicular1} (a). The measurements were performed in a $^3$He/$^4$He dilution refrigerator with a  base temperature of 20~mK using standard lock-in technique. Coulomb blockade measurements revealed the QD's charging energy of $U= 350\pm 30~\mu$eV. The number of electrons confined in the QD is estimated to be $61 \pm 3$.

Magnetic fields applied perpendicular to the 2DEG affect spin and orbital motion of the electrons leading to the formation of Landau levels in the QD and in the leads.  For a multi-electron QD a complicated level structure with many occupied Landau levels appears at small and intermediate magnetic fields. At large enough magnetic fields with only two Landau levels occupied the situation simplifies. For the Kondo effect only the outermost states of the QD belonging to the lowest Landau level (LL0) participate in Kondo transport \cite{Keller2001,Fuhner2002,Stopa2003}. An even or odd number of electrons on LL0 is determining whether Kondo enhanced transport is possible. Switching between even and odd occupation of LL0 is achieved either by redistribution of electrons between the two lowest Landau levels due to magnetic field modulation or by the loading of additional electrons to the dot by varying the gate voltage.  The resulting pattern is often called the \textit{Kondo chessboard}. A typical example is shown in Fig. \ref{perpendicular1} (c) for our sample.
The differential conductance $G= dI/dU_{sd}$ of the QD as function of perpendicular magnetic field $B_\perp$ and gate voltage is depicted there. At magnetic field values between 4~T and 5~T a regular pattern of alternating tiles of high and low conductance is visible. Low conductance corresponds to Coulomb blockade and high conductance to Kondo enhanced transport due to an odd number of electrons in LL0. The modulation caused by the magnetic field is shown in Fig. \ref{perpendicular1} (b) for a fixed gate voltage. At every step in the conductance a flux quantum is added to the QD and the electron number in LL0 is changed by one \cite{Fuhner2002}.

\begin{figure}
\includegraphics[scale=0.5]{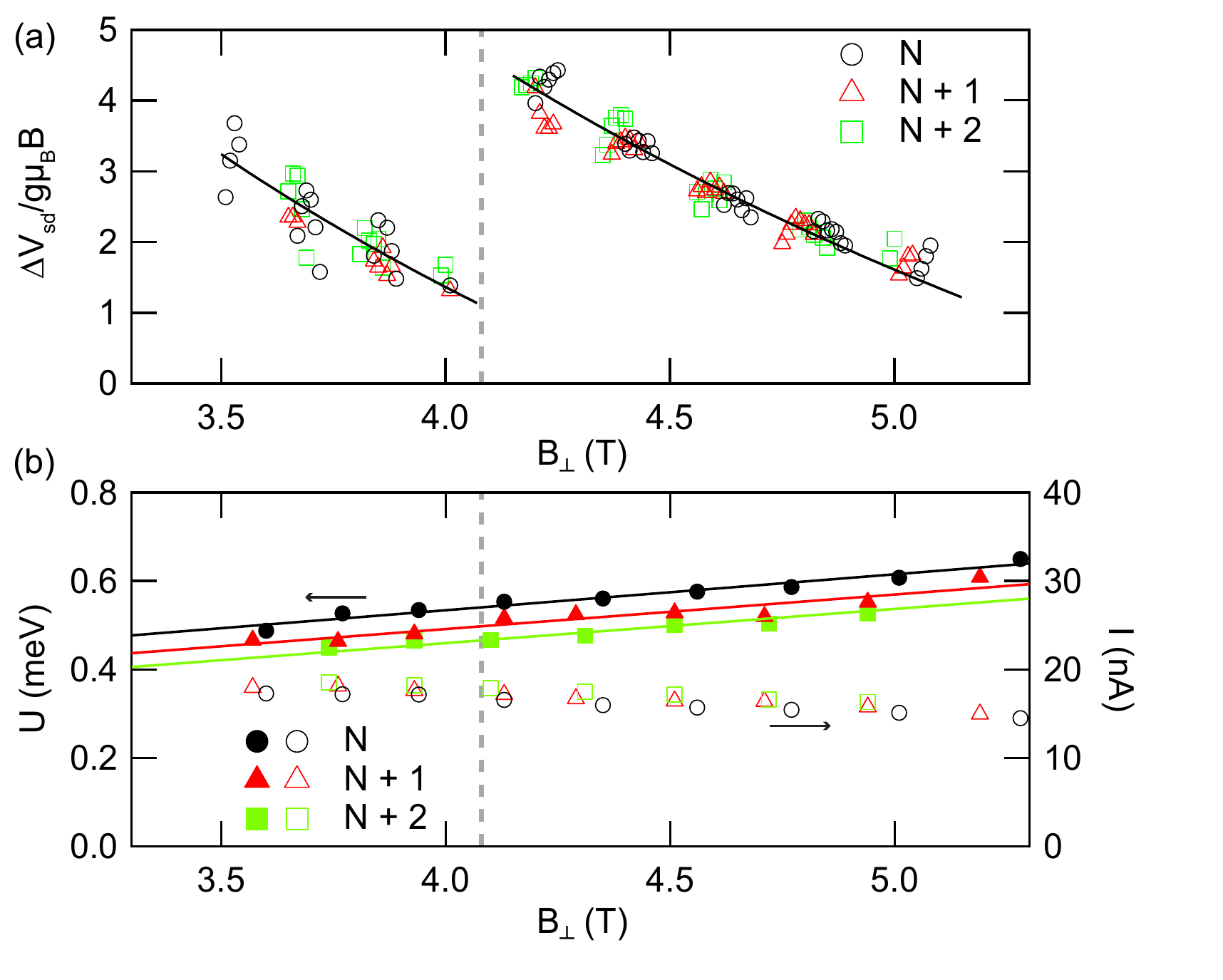}
\caption{(a) Evolution of the splitting width in the Kondo chessboard for three successive electron numbers in the range of $N\approx 60$. The dashed line indicates the integer filling factor $\nu=4$ in the leads. (b) Charging energy $U$ and current $I$ at fixed source drain voltage $V_{sd}=0.5$~mV in non-Kondo regions as function of magnetic field for the same electron numbers as shown in (a). Filled symbols show $U$ (left axis), open symbols show $I$ (right axis). The solid lines are a guide to the eye to show the linear increase of $U$.}
\label{perpendicular2}
\end{figure}

Measuring $G$ as function of bias in various regions of this \textit{Kondo chessboard} leads to the results seen in Fig. \ref{perpendicular1} (d-f). In tiles without Kondo transport (e), a minimum can be observed at zero-bias, as expected for a Coulomb blocked QD. In contrast a  higher conductance  at zero-bias and  two shoulders at finite bias can be distinguished in Fig. \ref{perpendicular1} (d) and (f) for the Kondo regime. These shoulders indicate a splitting of the ZBA. Comparing the splitting at 4.39~T (green triangle) and 4.57~T (black square) yields following observation: the splitting at 4.39~T, $\Delta V_{\text{sd}} = 0.38 \pm 0.02$~mV, clearly exceeds the one at 4.57~T, $\Delta V_{\text{sd}} = 0.32 \pm 0.01$~mV, a result which cannot be explained by bare Zeeman splitting $\Delta_Z=g\mu_BB$. To further examine the development of the splitting it was analyzed for three successive electron numbers, which is possible due to the chessboard pattern. A quantitative examination of the splitting width \cite{fitreference} in units of $\Delta_Z$ is shown in Fig. \ref{perpendicular2} (a). Quite astonishing the splitting decreases with increasing magnetic field and shows a discontinuity at $B_\perp=4.1$~T. In the following investigation of the underlying physics we focus on the discontinuity around $B=4.1$~T which we shall attribute to the quantum Hall effect (QHE) in the "metallic" leads. The considerations leading to this conclusion can be summarized as follows: first and foremost, the discontinuity does not depend upon the number of electrons on the QD. In fact, the values for the Zeeman splitting for different QD populations coincide which suggests that the discontinuity is not primarily associated with internal re-organization of the QD. Second, the discontinuity occurs at a magnetic field value where we anticipate an integer filling factor $\nu$ in the leads corresponding to two filled spin degenerate Landau levels. The step in the splitting width is much larger than expected for the influence of magnetic field modulated Fermi-energy on the Zeeman splitting due to non-parabolicity of the conductance band \cite{Dobers1988}.

Furthermore the step in the splitting width is not a result of a change in the coupling strength $\Gamma$ of the QD to the leads nor a change of the charging energy of the QD. The behaviour of $\Gamma$ which can be derived from the current $I$  through the QD decreases smoothly with increasing magnetic field as can be seen from Fig. \ref{perpendicular2} (b) for a fixed $V_{sd}=$0.5~mV. The evolution of $U$ was estimated by fitting the linear asymptotes of the current-voltage characteristic and determining the difference between these fits on both sides. The difference of the asymptotes is $U/2$ \cite{Bentum1988,Supplementary1}. Both $I$ and $U$ vary only slowly with magnetic field and exhibit no particular behaviour around 4.1 T as can be seen in Fig. \ref{perpendicular2} (b). The increase of $U$ and the decrease of $\Gamma$ with magnetic field are easily explained by compression of the electron wave functions on the QD and in the leads leading to a slightly lower capacitance i.e. a higher charging energy.

\begin{figure}
\includegraphics[scale=0.5]{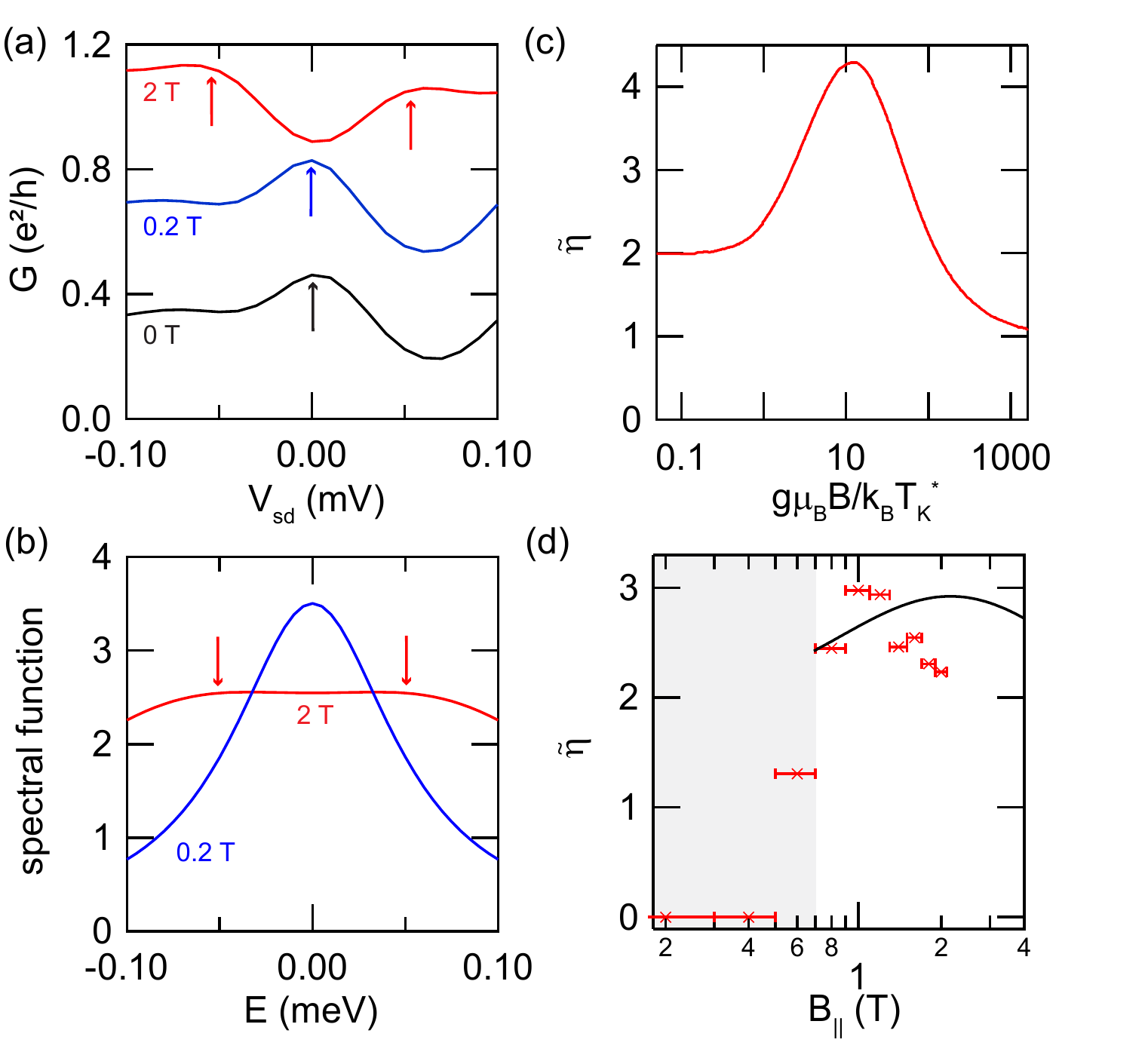}
\caption{(a) Differential conductance through the QD as function of bias. An offset of 0.35 and 0.7 $\text{e}^2/\text{h}$, respectively, is added to the traces at 0.2~T and 2~T for clarity.  The Kondo resonance at 0~T and 0.2~T and the symmetric peaks at 2~T are marked by arrows. (b) Calculation of the spectral function as function of energy for two different magnetic field values. The arrows point out the maxima of the function. (c) Enhancement factor $\tilde{\eta}$ over $g\mu_B B/k_BT_K^*$ on a logarithmic scale with $T_K^*=T_K/w$ being the Kondo temperature at $T=0$. The presented data was taken from Reference \cite{Hewson2006}. (d) Enhancement factor $\tilde{\eta}$  plotted as function of magnetic field. Red crosses show the experimental data. The black line represents results from renormalized perturbation calculations.}
\label{parallel1}
\end{figure}

Before we discuss the influence of the perpendicular magnetic field on the splitting width, we take a look at the Zeeman splitting of the ZBA in a parallel magnetic field. Application of magnetic fields in parallel  to the sample surface ($B_{||}$)  affects only the spins of the electrons due to the strong confinement of the 2DEG whereas application of a magnetic field perpendicular to the sample surface ($B_\perp$) and as such also perpendicular to the 2DEG influences also the orbital motion of the electrons.
Figure \ref{parallel1} (a) shows typical measurements of the differential conductance $G=dI/dV_{sd}$ as function of bias for different magnetic fields.  At $B=0$ the Kondo resonance at zero-bias is visible as a ZBA. The half width at half maximum  $\Delta_\text{ZBA}=50\pm 9~\mu$eV of this resonance is determined using a Lorentzian fit. The Kondo temperature $T_K = 188 \pm 34$~mK is estimated from $T_K=\pi w \Delta_\text{ZBA}/4 k_B$ with $w=0.4128$  being the Wilson number\cite{Hewson1993}.  The ZBA does not change for small values of $B_{||}$ as can be seen at $B_{||}=0.2$~T.  This insensitivity towards application of magnetic fields is expected as long as the width of the ZBA given by the Kondo temperature is larger than the Zeeman splitting.  For magnetic fields exceeding a threshold value of $B_\text{t} = 0.5\pm0.1$~T the peak starts to shift  indicating a Zeeman splitting of the resonance \cite{Kogan2004,Amasha2005,Quay2007}. Considering the g-factor for bulk GaAs $g=0.44$ our threshold value yields a bare Zeeman splitting of $g\mu_B B_\text{t} = 13\pm3\mu$eV which is of the same order of magnitude as the value for $\Delta_{ZBA}$ obtained at $B_{||}=0$.  Comparing the bare Zeeman splitting at the threshold to the width of the ZBA, however, we are lead to the conclusion that the actual splitting must be enhanced over its bare counterpart. A typical trace for the split Kondo peak is displayed for $B_{||}= 2$~T  in Fig. \ref{parallel1}(a). This can also be seen in Fig. \ref{parallel1} (b) where results for calculated spectral functions at 0.2~T and 2~T, respectively, are shown: at 0.2~T there is still only a single peak at zero-bias while at 2~T a sum of two Lorentz functions centred at $\pm$0.75~meV is obtained.

While the Anderson model for a correlated impurity with spin $S=1/2$ certainly has its limitations in describing the complex electronic structure of a QD with
$\sim$ 60 electrons, it nevertheless provides a semi-quantitative explanation of the observed phenomena. For simplicity we assume particle-hole
symmetry. In the limit $U\gg \pi \Gamma$, the model calculations shown in Fig. \ref{parallel1} (c) predict a non-monotonic variation of the enhancement factor $\tilde{\eta}=\tilde{g}/g$ from the low-field value $\tilde{\eta}=2$ to $\tilde{\eta}=1$ at high fields going through a maximum\cite{Hewson2006}. The existence of the latter can be easily understood from the competition between formation of a local moment and transversal spin fluctuations. This enhancement can be observed in the experimental results for $\Delta V_\text{sd}$ as obtained from the measurements that are shown in Fig. \ref{parallel1} (d). The quantitative deviations from the universal strong-coupling behaviour should be ascribed to the fact that the system under consideration is rather in the mixed-valent regime. In fact, the relatively large value $\Delta_{ZBA}/U$ suggests $U/\pi \Gamma=1.26$.
We adopt the Renormalized Perturbation Theory (RPT) \cite{Edwards2011} to calculate the variation with magnetic field of the enhanced Zeeman splitting \cite{Supplementary2}.
The results for $U=350~\mu$eV and $\Gamma=57~\mu$eV displayed in \ref{parallel1} (d) reproduce the experimental data rather well. The parameters chosen
underestimate the width of the Kondo resonance by a factor of 3. This discrepancy can be explained by the fact that we reduced the electronic structure of the QD to a single spin-degenerate level. It is well known from magnetic rare-earth impurities that including higher CEF states increases the prefactor.

Having explained the evolution with magnetic field $B_\parallel$ of the Kondo resonance we next turn to the fields $B_\perp$ applied perpendicular to
the QD. We do not attempt at a fully microscopic explanation for the discontinuous changes in the Zeeman splitting around 4.1~T but
rather present a qualitative picture for these novel phenomena. Assuming the different splitting width in the perpendicular field to be a consequence of a discontinuous variation of $\tilde{g}$ leads to a sharp change in $T_K$ since $B_\perp$ is quasi constant at this point. However, the change in $T_K$ is not the result of a change in the coupling between the QD and the leads or a variation of $U$ as shown above.
The key to an understanding of the findings above is that (i) the Kondo effect is essentially a screening process and that (ii) the screening properties of a 2DEG may depend sensitively on the filling factor of the Landau levels in the quantum Hall regime. To understand the interplay of these two phenomena, we have to compare their characteristic length scales.
The prevailing view of the Kondo effect is that at low temperatures $T \ll T_K$ the magnetic moment of the QD forms a singlet with the conduction electrons. The magnetic impurity becomes invisible outside a screening cloud of spatial extent $\xi_K=\hbar v_F/k_B T_K$, with $v_F$ being the Fermi velocity in a two-dimensional system and $\hbar$ the Planck constant. Since $k_BT_K$ is comparable to the width of the Kondo resonance $\Delta_{ZBA}$, one can write $\xi_K= \hbar v_F/\Delta_{ZBA}$. The importance of the Kondo scale has been confirmed only recently by scanning tunnelling spectroscopy techniques \cite{Pruser2011} and detailed theoretical studies \cite{Mitchell2011}.

The formation of the Kondo cloud reflects the screening properties of the leads. The latter, however, are known to change dramatically with magnetic field in the quantum Hall regime. It is generally accepted that the magnetic field acting on the orbital motion of the conduction electrons leads to the formation of compressible regions which are separated by incompressible strips. The characteristic length scale $l_B$ for the edge channels is set by electrostatics and is, as such, only weakly affected by external magnetic fields \cite{Chklovskii1992}. Simple estimates show that $l_B\approx 100$~nm $\ll \xi_K \approx 5000$~nm. As a consequence, the Kondo effect will be affected by changes in the edge structure but also by changes in the interior of the leads since both will introduce additional fine-structure in the local density of states close to the QD. Screening is possible only from the compressible metallic regions. As a result, we anticipate a (partial) suppression of the Kondo effect when large parts of the leads become incompressible. This suppression happens indeed whenever the bulk filling factor approaches an integer value from above \cite{Zhitenev1993} as is illustrated in Fig. \ref{edgechannels}. In this field regime compressible regions exist only close to the edges and  screening originates almost exclusively from edge channels. For filling factors slightly below an integer value, on the other hand, large portions of the leads are compressible and should be available for screening leading to a rise in the Kondo temperature.

It is tempting to estimate the changes in $T_K$ and concomitantly of the Kondo coupling from the observed discontinuity in the Zeeman splitting at integer filling factor. Thereby the effective splitting width was assigned to a corresponding $B/T_K$ via the results in Fig. \ref{parallel1} (c) and $T_K$ was calculated from the known applied magnetic field values. The results are Kondo temperatures between  $T_K=9\pm5$~mK at $B_\perp = 4$~T and $T_K= 69\pm32$~mK at 4.2~T.  A $T_K$ of $69$~mK is roughly a factor of 2 smaller than  $T_K$ at $B = 0$. This reduction in comparison to the field-free value is  understood in taking into account  the reduced coupling between the QD and the leads and the small incompressible regions of the leads.  Kondo temperatures of only $9\pm5$~mK as computed at  $T_K$ at $B_\perp = 4$~T are of the same order or even smaller than our refrigerator's base temperature of 20~mK. In this regime the properties of the system are non-universal and not really described by the theory shown in Fig. \ref{parallel1} (c). However, Kondo enhanced conductance in QDs can still be visible at temperatures above $T_K$ \cite{Goldhaber-Gordon1998}. At $B_\perp = 4$~T the corresponding shoulders in $G$ are barely visible, which implies a Kondo temperature lower than our electron temperature. Therefore, the extracted Kondo temperatures are still in agreement with the explanation for the development of $\Delta V_\text{sd}$ using Fig. \ref{parallel1} (c).  The small $T_K$ at $B_\perp = 4$~T is  a consequence of the reduced screening by the conduction electrons due to extended incompressible regions in the leads.

In summary, we measured the Zeeman splitting of the ZBA in parallel and perpendicular magnetic field. While in parallel fields the splitting increases non-monotonically with magnetic field  as expected from the renormalized g-factor, a noticeable observation was made in the perpendicular field case. In the so-called Kondo chessboard the splitting of the Kondo resonance decreases with increasing magnetic field and shows a sharp discontinuity where the filling factor in the two-dimensional leads reaches an integer value. This behaviour shows the dependence of the renormalized Zeeman splitting of the Kondo resonance on the screening properties of the two-dimensional leads which are formed by compressible states contributing to screening and incompressible states which cannot participate in Kondo screening.

\begin{acknowledgements}
We thank B. Popescu for producing the sample and thank B. Shklovskii and A.C. Hewson for discussions. We acknowledge funding by the  School for Contacts in Nanosystems.
\end{acknowledgements}



\begin{thebibliography}{10}

\bibitem{Hewson1993}
A.~C. Hewson, {\em The Kondo Problem to Heavy Fermions}. \newblock (Cambridge University Press, Cambridge, UK, 1993).

\bibitem{Goldhaber-Gordon1998a}
D. Goldhaber-Gordon, H. Shtrikman, D. Mahalu, D. Abusch-Magder, U. Meirav, and M.~A. Kastner, Nature \textbf{391}, 156 (1998).

\bibitem{Goldhaber-Gordon1998}
D. Goldhaber-Gordon, J. G\"{o}res, M.~A. Kastner, H. Shtrikman, D. Mahalu, and U. Meirav, Phys. Rev. Lett. \textbf{81}, 5225 (1998).

\bibitem{Cronenwett1998}
S.~M. Cronenwett,  T.~H. Oosterkamp, and L.~P. Kouwenhoven, Science \textbf{281}, 540 (1998).

\bibitem{Schmid1998}
J. Schmid, J. Weis, K. Eberl, and K. v. Klitzing, Physica B \textbf{256--258}, 182 (1998).

\bibitem{Wiel2000}
W.~G. van~der Wiel, S.~De Franceschi, T. Fujisawa, J.~M. Elzerman, S. Tarucha, and L.~P. Kouwenhoven, Science \textbf{289}, 2105 (2000).

\bibitem{Mourik2012}
V. Mourik, K. Zou, S.~M. Frolov, S.~R. Plissard, E.~P.~A.~M. Bakkers, and L.~P. Kouwenhoven, Science \textbf{336}, 1003 (2012).

\bibitem{Lutchyn2010}
R.~M. Lutchyn, J.~D. Sau, and S. Das Sarma, Phys. Rev. Lett. \textbf{105}, 077001 (2010).

\bibitem{Oreg2010}
Y. Oreg, G. Refael, and F. von Oppen, Phys. Rev. Lett. \textbf{105}, 177002 (2010).


\bibitem{Hewson2006}
A.~C. Hewson, J. Bauer, and W. Koller, Phys. Rev. B \textbf{73}, 045117 (2006).

\bibitem{Edwards2011}
K. Edwards,  and A.~C. Hewson, J. Phys.: Condens. Matter \textbf{23}, 045601 (2011).

\bibitem{Wilson1975}
K.~G. Wilson, Rev. Mod. Phys. \textbf{47}, 773 (1975).

\bibitem{Held1998}
R. Held, T. Vancura, T. Heinzel, K. Ensslin, M. Holland, and W. Wegscheider, Appl. Phys. Lett. \textbf{73}, 262 (1998).

\bibitem{Keyser2000}
U.~F. Keyser, H.~W. Schumacher, U. Zeitler, R.~J. Haug, and K. Eberl, Appl. Phys. Lett. \textbf{76}, 457 (2000).

\bibitem{Tutuc2011}
D. Tutuc, B. Popescu, D. Schuh, W. Wegscheider, and R.~J. Haug, Phys. Rev. B \textbf{83}, 241308 (2011).

\bibitem{Keller2001}
M. Keller, U. Wilhelm, J. Schmid, J. Weis, K.~v. Klitzing, and K. Eberl, Phys. Rev. B \textbf{64}, 033302 (2001).

\bibitem{Fuhner2002}
C. F\"{u}hner, U.~F. Keyser, R.~J. Haug, D. Reuter, and A.~D. Wieck, Phys. Rev. B \textbf{66}, 161305 (2002).

\bibitem{Stopa2003}
M. Stopa, W.~G. van~der Wiel, S.~De Franceschi, S. Tarucha, and L.~P. Kouwenhoven, Phys. Rev. Lett. \textbf{91}, 046601 (2003).

\bibitem{fitreference}
To determine the position of the shoulders in $G$, they were fitted with a Gaussian function and an additional cubic background.


\bibitem{Dobers1988}
M. Dobers, K.~v. Klitzing, and G. Weimann, Phys. Rev. B \textbf{38}, 5453 (1988).

\bibitem{Bentum1988}
P.~J.~M. van Bentum, H. van Kempen, L.~E.~C. van de Leemput, and P.~A.~A. Teunissen, Phys. Rev. Lett. \textbf{60}, 369 (1988).

\bibitem{Supplementary1}
More details on the extraction of $U$ can be found in the supplementary material.

\bibitem{Kogan2004}
A. Kogan, S. Amasha, D. Goldhaber-Gordon, G. Granger, M.~A. Kastner, and H. Shtrikman, Phys. Rev. Lett. \textbf{93}, 166602 (2004).

\bibitem{Amasha2005}
S. Amasha, I.~J. Gelfand, M.~A. Kastner, and A. Kogan, Phys. Rev. B \textbf{72}, 045308 (2005).

\bibitem{Quay2007}
C.~H.~L. Quay, J. Cumings, S.~J. Gamble, R. de Piciotto, H. Kataura, and D. Goldhaber-Gordon, Phys. Rev. B \textbf{76}, 245311 (2007).

\bibitem{Supplementary2}
A description of this method is provided in the supplementary material.


\bibitem{Pruser2011}
H. Pr\"{u}ser, M. Wenderoth, P.~E. Dargel, A. Weismann, R. Peters, T. Pruschke,  and R.~G. Ulbrich, Nature Phys. \textbf{7}, 203, (2011).

\bibitem{Mitchell2011}
A.~K. Mitchell, M. Becker,  and R. Bulla, Phys. Rev. B \textbf{84}, 115120 (2011).

\bibitem{Chklovskii1992}
D.~B. Chklovskii, B.~I. Shklovskii,  and L.~I. Glazman, Phys. Rev. B \textbf{46}, 4026 (1992).


\bibitem{Zhitenev1993}
N.~B. Zhitenev, R.~J. Haug, K. v. Klitzing, and K. Eberl, Phys. Rev. Lett. \textbf{71}, 2292 (1993).

\end{thebibliography}
\end{document}